\documentclass[pre,onecolumn,showpacs,preprint]{revtex4}

\usepackage[dvips]{graphicx}

\begin{document}

\preprint{}

\title{Topological properties of the contact network of granular packings.}

\author{Roberto Ar\'{e}valo}
\affiliation{Departamento de F\'{\i}sica, Facultad de Ciencias, Universidad de Navarra,
E-31080 Pamplona, Spain.}
\author{Iker Zuriguel}\affiliation{Departamento de F\'{\i}sica, Facultad de Ciencias, Universidad de Navarra,
E-31080 Pamplona, Spain.}
\author{Diego Maza}
\affiliation{Departamento de F\'{\i}sica, Facultad de Ciencias, Universidad de Navarra,
E-31080 Pamplona, Spain.}


\begin{abstract}
The force networks of different granular ensembles are defined and
their topological properties studied using the tools of complex
networks. In particular, for each set of grains compressed in a
square box, it is introduced a force threshold that determines
which contacts conform the network. Hence, the topological
characteristics of the network are analyzed as a function of this
parameter. The characterization of the structural features thus
obtained, may be useful in the understanding of the macroscopic
physical behavior exhibited by this class of media.

\end{abstract}

\volumeyear{}
\volumenumber{}
\issuenumber{}
\eid{}
\date{24 May 2007}
\startpage{1}
\endpage{}
\maketitle

\section{Introduction}

Granular materials are being widely studied by the physics
community since they exhibit unusual and distinctive properties
\cite{duran}. These materials are composed of macroscopic
particles that interact by a dissipative contact force and can be
thought of as displaying gas, liquid and solid phases. A suitable
model for the study of granular materials is to consider each
grain as a hard sphere, ignoring fragmentation and moving the
effect of deformation to the dissipative term. As pointed in
\cite{anikeenko}, the wide applicability of these model to the
study of liquids, glasses and colloids implies a paramount
importance of the geometrical properties of the packing of hard
spheres in determining the physics exhibited by the materials
analyzed. The geometry of granular packing has been investigated
\emph{i.e.} by  \cite{anikeenko} and \cite{aste} using
Voronoi-Delaunay partitioning to identify structures in the former
and volume distributions in the latter.

\begin{figure}
\center
\includegraphics[width=0.4\textwidth, keepaspectratio]{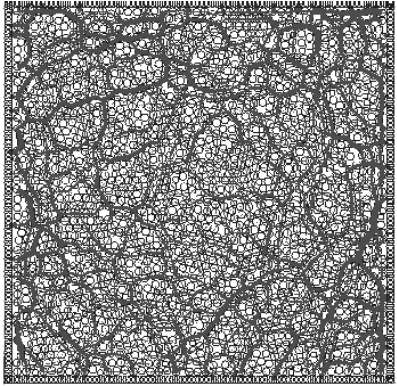}
\caption{Example of a compressed sample obtained with our
simulations showing the force chains. The thickness of the lines
represents the magnitude of the normal force originated by the
interaction between each pair of particles.} \label{fig1}
\end{figure}

In the present work we propose, in the same line of those and
other works \cite{ostojic}, a structural study of granular packing
but using tools specifically developed in the frame of complex
networks. As will be explained later we define for each packing a
network of contacts, see FIG.~\ref{fig1}, which topological
properties are studied afterwards.

The contact topology of a granular packing can be studied as a
graph where particles are \emph{nodes} and the interacting force
pairs \emph{edges}. This approach has important advantages. For
one hand, it is a quantitative tool as are not other ideas
proposed in the granular community, namely, that of ``force
chains" \cite{peters}; it is an abstract point of view that allows
to reach very primitive concepts such as connectivity over which
to elaborate more complex definitions; and, finally, the field of
complex networks provides us with a great amount of concepts and
algorithms among which we can chose the most suitable for our
proposes of characterization.

The remainder of this work is structured as follows: in section
$2$ we explain the numerical method used and the protocol followed
to obtain the samples that we study. In section $3$ the
topological properties analyzed are defined and results presented
for several different conditions. Finally, in section $4$ we
summarize our results and draw some conclusions.

\section{Numerical method}

We perform soft particle molecular dynamics simulations of discs
in two dimensions. The model of contact includes a linear
restoring force in the normal direction of the impact and a
tangential force providing static friction. The complete details
of the simulation protocol are carefully described in
\cite{arevalo}. The values used for the parameters of the force
model are: the frictional coefficient ($\mu=0.5$), the elastic
constant ( $k_{n}=10^{5}$), a dissipative coefficient $\gamma_{n}=150$, 
and the corresponding ones for the tangential component (
$k_{s}=\frac{2}{7}k_{n}$ and $\gamma _{s}=300$)  with an
integration time step $\delta =10^{-4}\tau$. The stiffness
constants $k$ are measured in units of $mg/d$, the damping
constants $\gamma$ in $m\sqrt{g/d}$ and time in $\sqrt{d/g}$.
Here, $m$, $d$ and $g$ stand, respectively, for the mass of the
discs, the diameter of the discs and the acceleration of gravity.

A typical simulation starts by randomly placing the discs in a
wide horizontal area such that no one of them is in contact with
any other and the packing fraction is around $0.1$. Discs are
given random velocities drawn from a gaussian distribution. Four
walls made up with the same grains that constitute the bulk
compress the system until a certain predefined threshold of force
is attained. It is important to note that due to the dissipative
nature of the interactions the final kinetic energy is vanishing
small. The final configuration obtained is saved in order to be
carefully analysed. This configuration is named a ``jammed state"
by the granular community and essentially corresponds to a
metastable equilibrium state compatible with the history of the
configuration.

We run simulations under several different conditions to check the
variation of the results with the number of grains,
polydispersity, friction coefficient, maximum applied pressure and
geometry of the compression cell. Let us call sample A that
obtained with a bidisperse mixture of discs, $15\%$ with radii $d$
and the rest with radii $7/9d$ and parameter values as given
above; sample B has the same properties than A but the disks are
monodisperse with radius $d$; sample C is bidisperse as A but the
friction coefficient is $\mu=0.25$; sample D is the same as C but
the final pressure is increased a $50\%$; finally, sample E is the
same as A but the bounadary conditions are circular instead of
square. For samples A and B we run simulations with $512$, $1024$
and $2048$ discs, for samples C, D and E only with $2048$ since,
as will be shown, no significant dependence on system size is
found. In order to attain good statistics we perform $20$
independent simulations for each sample and average the results.

\section{Contact network as a complex network}

In the first place we define our network, i.e. a set of nodes
connected by edges \cite{newman}, as follows. Every grain with, at
least, one contact constitutes a node and edges are the
connections between the grains (nodes) in contact. There is a
contact between two grains if the distance between the centre of
them is smaller or equal to its diameter. A contact defines a
certain amount of normal force $F$ between the grains. Such
situation has been deeply studied by many authors using mainly
lattices diffusive models \cite{liu}. One of the main results of
these works is that, independently of the system details, the mean
value of the force distribution $\!\!<\!\!F\!\!>$ is a typical
scale of the problem. Nevertheless, many open questions remain
open about the properties of these systems: why the fluctuations
in the force distribution are as large as $\;\!\!<\!\!F\!\!>$ ?
Which statistical framework is suitable to explain the
experimental results?

In order to study the role of the topology on this problem we will
use the tools introduced in the theory of complex networks. We
introduce a force threshold $f$ such that any contact with a
normal force bigger than $f$ is an edge, but contacts with lower
values of normal force are not edges and grains with no contact
are not nodes.  Thus we obtain a network which depends on $f$, and
hence its topological properties can be studied as a function of
$f$. In our definitions we do not consider the grains of the
walls. In the remainder of this section we present the results
obtained for each of the topological properties studied along with
their definitions \cite{newman, costa}.

\subsection{Connectivity}

\begin{figure}
\includegraphics[width=0.8\textwidth,keepaspectratio]{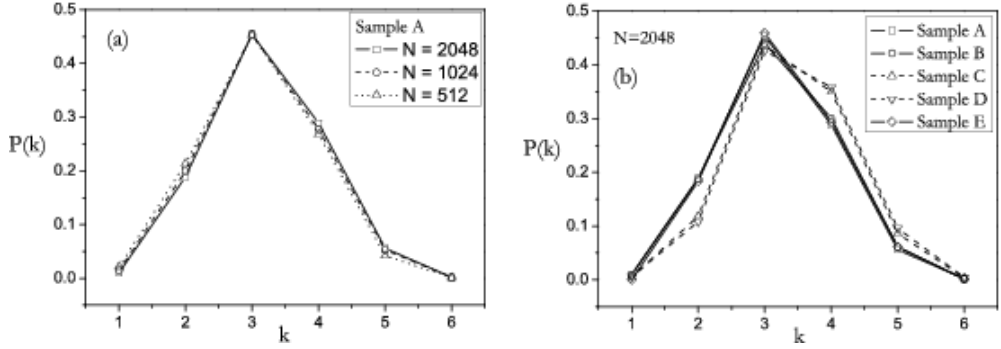}
\caption{Left panel shows $P\left(k\right)$ for sample A using
$N=2048,1024,512$ discs. In the right panel we show
$P\left(k\right)$ with $N=2048$ for all samples.} \label{fig2}
\end{figure}

In our case, the connectivity $k$ of a node represents the number
of contacts between  neighboring particles. Then, the degree
distribution $P\left(k\right)$ is the distribution function of the
number of contacts per particle. In FIG.~\ref{fig2}a we show the
degree distribution of sample A for three different sample sizes
showing that there is no substantial variation. In all the cases
the maximum $P\left(k\right)$ is found for $k=3$ and around $95\%$
of the particles present values of $k$ between 2 and 4. The degree
distribution for the rest of the samples with $N=2048$ is shown in
FIG.~\ref{fig2}b. The overall behavior of the function
$P\left(k\right)$ remains the same for all samples and only slight
deviations are appreciated for samples C and D. In particular,
samples C and D display higher number of nodes with higher values
of $k$. This result can be understood if it is considered that
samples C and D are the ones with the smaller friction
coefficient. This will result in a small amount of arches inside
the sample and consequently a reduction of the amount of particles
that display just two contacts.

\begin{figure}
\includegraphics[width=0.8\textwidth, keepaspectratio]{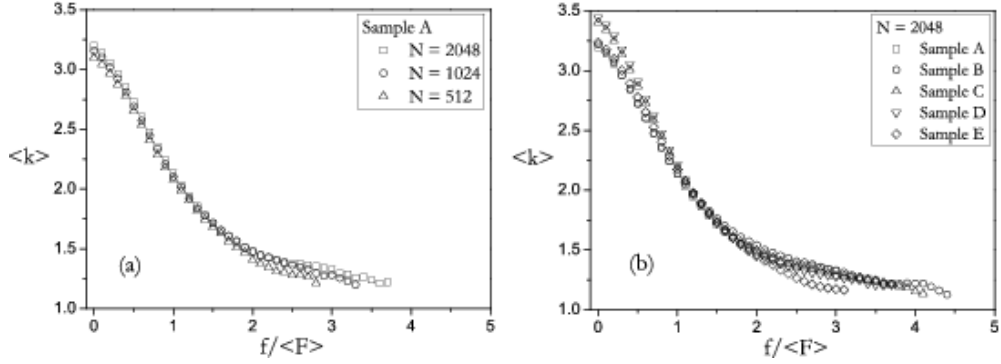}
\caption{Left panel: average connectivity of sample A. Right
panel: average connectivity of all samples.} \label{fig3}
\end{figure}

In figure FIG.~\ref{fig3}a we show the behavior of the average
connectivity $<k>$ as a function of the force threshold
$f/\!\!<\!\!F\!\!>$ for sample A. Again, this property seems
independent of the system size. The figure FIG.~\ref{fig3}b shows
the results obtained for the different samples. Again, small
differences are appreciable for samples C and D without
modification of the overall behavior.

The most prominent feature is a fast decay of the connectivity
upon increasing the force threshold. It could be said that the
small forces are the ones which keep the network connected and the
connectivity almost disappears when they are removed.

\subsection{Geodesic distance and network diameter}
\begin{figure}

\includegraphics[width=0.8\textwidth, keepaspectratio]{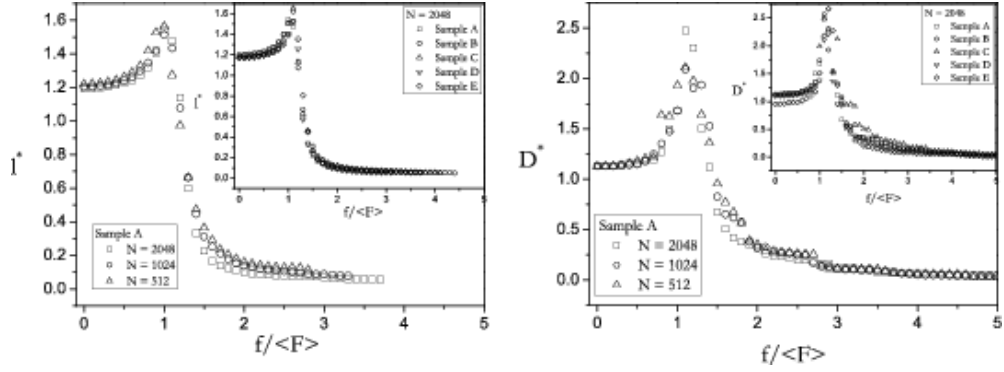}
\caption{Left panel:$l^*$ is the geodesic distance $l$ normalized
by $\sqrt{N}/2$ for sample A. The inset shows $l^*$ for the all
the samples with $N=2048$. Right panel: $D^*$ is the network
diameter $D$ normalized by $\sqrt{2N}$ for sample A. The inset
shows $D^*$ for all the samples with $N=2048$.} \label{fig4}
\end{figure}

The geodesic distance $l$ between two nodes is the smallest number
of edges that separate them. This quantity can be measured by a
number of algorithms, we used the \emph{breadth first search}. The
diameter $D$ of the network is the longest of the geodesic
distances. In figure FIG.~\ref{fig4} the normalized geodesic
distance $l^*$ as a function of $f/\!\!<\!\!F\!\!>$ is shown for
sample A. The geodesic distance is normalized by $\sqrt{N}/2$
since the geodesic distance increases with the number of particles
conforming the sample $N$. This scaling of $l$ with the network
size is what would correspond to a square lattice, so in the limit
case of $f/\!\!<\!\!F\!\!>\rightarrow0$ our network seems to be
not very different of a square one. The peak near
$f/\!\!<\!\!F\!\!>=1$ can be explained in terms of the polygons
that appear in the network as will be shown later. In the inset of
FIG.~\ref{fig4} it is shown that there is no difference in the
behavior of the geodesic distance for the different samples .

The right panel of  FIG.~\ref{fig4} shows $D^*$, the network
diameter normalized by $\sqrt{2N}$. This normalization factor is
applied to show that the diameter of the network scales with the
diagonal of the compression cell. The behavior is entirely similar
to that of the geodesic distance and only a small deviation is
noticeable for sample E which was generated with a circular cell.
For this reason we can attribute this minor difference to the
scaling factor.

\subsection{Number of nodes and maximum cluster size}

\begin{figure}

\includegraphics[width=0.8\textwidth, keepaspectratio]{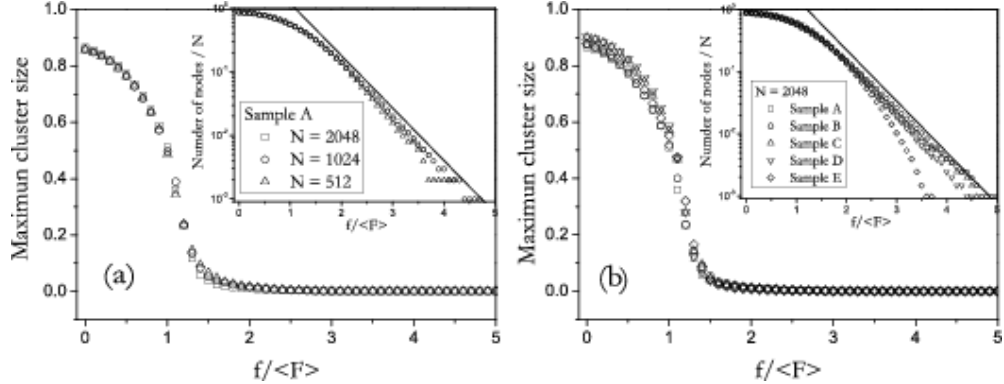}
\caption{(a) Size, in number of nodes normalized by $N$, of the
largest cluster in the network for sample A. Inset: Total number
of nodes in the network normalized by $N$ for sample A with
$N=512, 1024, 2048$. (b) Size, in number of nodes normalized by
$N$ for all the samples and $N=2048$. Inset: total number of nodes
for all the samples and $N=2048$. The line in both graphics has
slope $1.9$.} \label{fig5}
\end{figure}

We define a cluster as a group of nodes mutually connected. The
total number of nodes in the network includes nodes from different
clusters. In the inset of FIG.~\ref{fig5}.a the total number of
nodes is presented for sample A normalized by $N$, in
semilogarithmic scale, showing that no variation appears upon
increasing the network size. The inset of FIG.~\ref{fig5}.b shows
the result for the rest of the samples. The number of nodes in the
network decays exponentially, the line in both figures has slope
$1.9$, as the force threshold $f/\!\!<\!\!F\!\!>$ is increased
beyond $f/\!\!<\!\!F\!\!>=1$. Before the point
$f/\!\!<\!\!F\!\!>=1$ the number of nodes decays only slightly.
For $f/\!\!<\!\!F\!\!>=0$ the normalized number of nodes is not
$1$ implying that there are grains without any contact. This is
due to the frictional nature of the medium which is able to create
arches surrounding one or more grains.

As $f/\!\!<\!\!F\!\!>$ increases the network disaggregates in
clusters that are not connected to each other. In
FIG.~\ref{fig5}.a the size of the largest cluster, measured in
number of grains, is shown for sample A normalized by $N$ while
FIG.~\ref{fig5}.b shows this quantity for the rest of the samples.
The largest cluster size dramatically drops in the vicinity of
$f/\!\!<\!\!F\!\!>=1$ and is almost zero beyond
$f/\!\!<\!\!F\!\!> \simeq 1.5$.

\subsection{Properties of clusters}

\begin{figure}
\center
\includegraphics[width=0.5\textwidth, keepaspectratio]{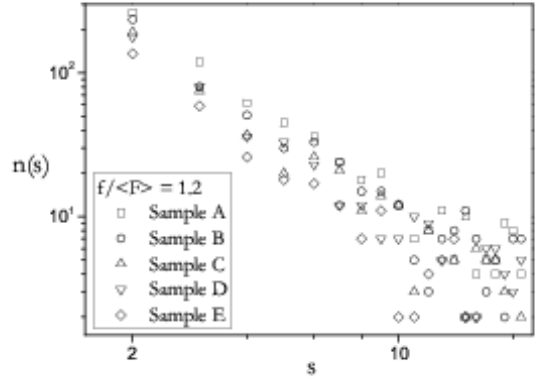}
\caption{The distribution of cluster sizes $s$ for all samples at
$f/\!\!<\!\!F\!\!>=1.2$ showing that it is a power function.}
\label{fig6}
\end{figure}

In this section we further analyze the properties of clusters as
defined in the previous section. As it is done in percolation
theory \cite{ostojic} we remove the largest cluster, which has yet
been analyzed, and study the distribution of the sizes $s$ of the
remaining clusters for different values of the force threshold:
$n\left(s,f/\!\!<\!\!F\!\!>\right)$. We use the samples with
$2048$. In figure FIG.~\ref{fig6} $n\left(s,1.2\right)$, the
cluster size distribution for $f/\!\!<\!\!F\!\!>=1.2$, is shown
for all samples. In logarithmic scale it can be fitted by a line
whose slope, in this case is around $1.9$. We have enough
statistics only for values of $f/\!\!<\!\!F\!\!>$ between $1$ and
$3$ and in this range we find that the distribution of sizes
behaves like $n\left(s,f/\!\!<\!\!F\!\!> \right) \propto
s^{\alpha}$ with $\alpha$ varying with $f/\!\!<\!\!F\!\!>$.

\begin{figure}
\center
\includegraphics[width=0.5\textwidth, keepaspectratio]{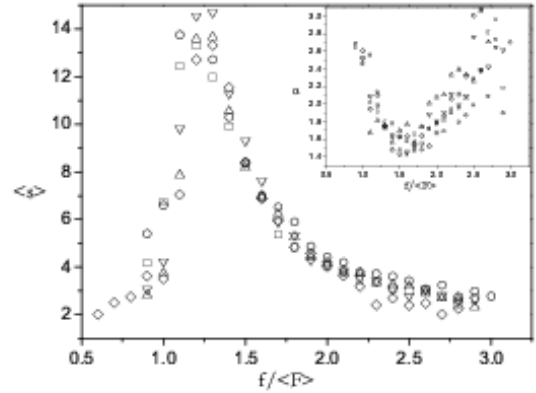}
\caption{Average cluster size, removing the largest one, as a
function of the force threshold $f/\!\!<\!\!F\!\!>$. The inset
shows the power $\alpha$ of the cluster size distribution
$n\left(s,f/\!\!<\!\!F\!\!>\right)$ as a function of
$f/\!\!<\!\!F\!\!>$.} \label{fig7}
\end{figure}

In FIG.~\ref{fig7} we show the average cluster size $<s>$ as a
function of $f/\!\!<\!\!F\!\!>$ with an inset showing the behavior
of $\alpha$. If we had retained the largest cluster to compute the
sizes distribution, FIG.~\ref{fig7} would display a monotonically
increasing function upon decreasing $f/\!\!<\!\!F\!\!>$. Instead,
it reveals a characteristic feature, a peak around
$f/\!\!<\!\!F\!\!>\simeq 1.2$ which is accompanied by a minimum in
$\alpha\left(f/\!\!<\!\!F\!\!>\right)$ around
$f/\!\!<\!\!F\!\!>=1.5$.

\subsection{Fractal dimension}

\begin{figure}
\includegraphics[width=0.8\textwidth, keepaspectratio]{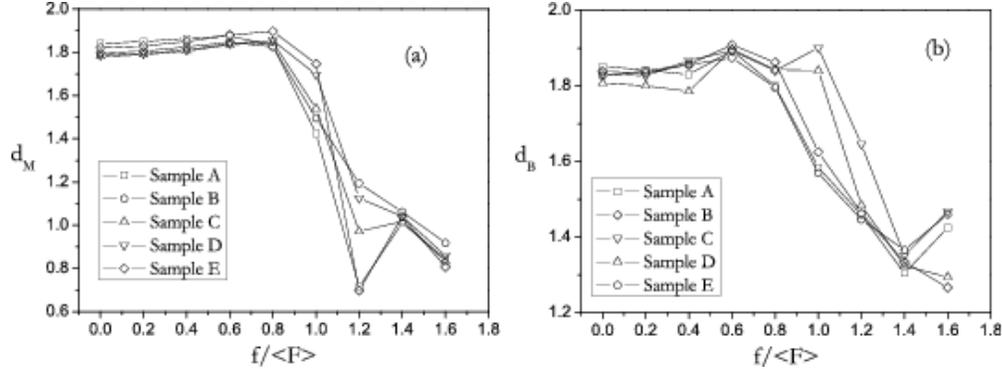}
\caption{Left panel: the mass fractal dimension as a function of
$f/\!\!<\!\!F\!\!>$ for all samples with $N=2048$. Right panel:
the box counting fractal dimension as a function of
$f/\!\!<\!\!F\!\!>$ for all samples with $N=2048$.} \label{fig8}
\end{figure}

In the theory of critical phenomena the value of the fractal
dimension determines the universality of a system, and thus, a set
of properties. In this section we compute the fractal dimension as
a function of $f/\!\!<\!\!F\!\!>$. Two such fractal dimensions can
be defined \cite{song1}: the mass fractal dimension and the box
counting fractal dimension. The former is computed choosing a node
and tracing circumferences of increasing radius $R$ around it. The
mass $M$, in number of nodes, inside each circumference is
computed and if it behaves like $M\propto R^{d_{M}}$ then $d_{M}$
is the mass fractal dimension. This procedure is repeated changing
the initial node and averaging the results. The boxcounting
fractal dimension is computed analyzing how the minimum $N_B$
number of boxes necessary to cover the network changes with the
box size $L$. If this verifies $N_B\propto L^{d_B}$ then $d_B$ is
the box counting fractal dimension. The process of minimization
involved in the last calculation renders it non immediate and we
followed the methods exposed in \cite{song2}. The results obtained
for both, the mass fractal dimension and the box counting fractal
dimension, are shown in FIG.~\ref{fig8} as a function of the force
threshold $f/\!\!<\!\!F\!\!>$.

Both dimensions are fairly equal to $1.8$ for values of
$f/\!\!<\!\!F\!\!>$, roughly, lower than $1$. A slight increase
can be perceived from $f/\!\!<\!\!F\!\!>=0$ until
$f/\!\!<\!\!F\!\!>\simeq 1$ where a marked drop takes place. This
fall of the fractal dimension is sharper and deeper for the mass
dimension but clearly present in both cases. The calculation of
the fractal dimensions cannot be carried out beyond the limit
shown since the network rapidly disaggregates. Thus we find a
change of behavior of the contact network in the vicinity of
$f/\!\!<\!\!F\!\!>=1$ that could be assigned to a change in the
universality class that describes the network as a function of the
force threshold.

\subsection{Third order loops of contacts}

A third order loop is defined as a three-step walk whose first and
last nodes are the same. Third order loops are thus contacts
arranged in a triangular fashion whose number can be computed by
the clustering coefficient \cite{newman,costa} or the third moment
of the adjacency matrix \cite{goh}. In rigidity theory
\cite{rigidity} these are, in two dimensions, the simplest rigid
structures. Indeed if we think in a triangle whose edges are rigid
and joined by freely rotational hinges it remains undeformed upon
external perturbations. On the contrary a square made of rigid
edges and freely rotational hinges is easily deformed by shear in
parallel sides. It is important to note that it is a sufficient,
but not necessary, condition for a polygon to be rigid that all
its faces are composed of triangles. Thus, triangles may be
important for the rigidity displayed by granular packings in two
dimensions.

\begin{figure}
\center
\includegraphics[width=0.4\textwidth, keepaspectratio]{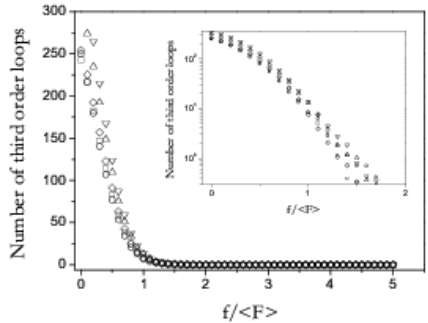}
\caption{Number of third order loops as a function of
$f/\!\!<\!\!F\!\!>$ for all samples with $N=2048$. The inset shows
the same data in semilogarithmic scale.} \label{fig9}
\end{figure}

In  FIG.~\ref{fig9} the number of triangles is reported for all
samples. There are no triangles beyond $f/\!\!<\!\!F\!\!>\geq 1.5$
and they concentrate in the edges carrying a small amount of
normal force, increasing exponentially when decreasing the force
threshold. We believe that the apparition of third order loops of
contacts is at the heart of the behavior found for some of the
topological properties presented in this paper as the geodesic
distance and the network diameter. For force values above
$f/\!\!<\!\!F\!\!>= 1.5$ a decrease in the force threshold
provokes the connection of different clusters of the network and
then, both the geodesic distance and the network diameter grow.
However, a further decrease in the force threshold below
$f/\!\!<\!\!F\!\!>= 1.5$ implies the apparition of third order
loops which will reduce the geodesic distance between the nodes
that belong to them, and hence the diameter of the network. The
way in which a third order loop reduces the geodesic distance
between nodes can be easily understood. If we imagine three nodes
(a,b,c) where the connections are a-b and b-c, the mean geodesic
distance in the cluster will be 1.33 as a-b and b-c are separated
by one edge but a-c are separated by two edges. In the same way
the diameter of this small network will be $2$ as it is the
maximum geodesic distance between the nodes. If now the nodes a
and c are also connected giving rise to a third order loop, both
the mean geodesic distance and the diameter of the network will be
reduced to 1.

\section{Discussion}

In this work we report on some topological properties of the force
interactions of granular packings by means of ideas specifically
introduced for complex networks. Our aim is to characterize the
heterogeneity of these systems without using definitions that may
change from one author to another, like usually occurs in the case
of force chains.

\begin{figure}
\center
\includegraphics[width=0.4\textwidth, keepaspectratio]{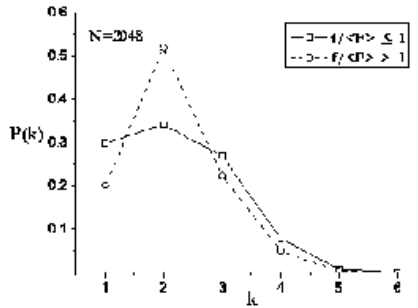}
\caption{Connectivities distribution for sample A where the nodes
connected by a force $f/\!\!<\!\!F\!\!> > 1$ and
$f/\!\!<\!\!F\!\!>\leq 1$ have been treated separately. The
subnetworks thus obtained exhibit well differentiated behaviors in
the region of low connectivity.} \label{fig10}
\end{figure}

The set of properties analyzed is insensitive to the size of the
system and shows only slight variations of behavior when the
friction coefficient or the applied pressure are changed. Thus,
they constitute a robust and useful description of an
heterogeneous material like the packing studied. It is noteworthy
that all the properties that have been analyzed as a function of
the force threshold $f/\!\!<\!\!F\!\!>$, that determines if an
edge is present in the network or not, display some distinctive
feature or sharp variation in the vicinity of
$f/\!\!<\!\!F\!\!>\simeq 1$. This behavior is indicative of a
change in the structural properties of the network in this point.
As has been pointed out by Radjai \emph{et al.} there seems to be
two subnetworks in the network of contacts, one ``weak" network
composed of small forces and containing around $60\%$ of grains,
and a ``strong" network constituted by edges carrying a force
above the average.

Our findings seem to support the existence of these two
subnetworks. As the force threshold is increased we remove the
weak network and retain only the strong one, leading to dramatic
changes that signal the change of behavior expected if both
subnetworks exist and are intrinsically different. In
FIG.~\ref{fig10} we show the connectivities distribution for both
subnetworks. It can be checked that they are quite similar for
high connectivities while differ significatively around $k=2$.

A natural extension of this work is to consider the intensity of
the force in every edge of the network and define weighted
networks. This point of view could be a more suitable tool in
order to relate structural features of the network with the
physical properties of the packing; in particular, it could shed
light into the question of the change of behavior at
$f/\!\!<\!\!F\!\!>\simeq 1$.

\section*{Acknowledgements}

This work has been supported by project FIS2005-03881 (MEC,
Spain), and PIUNA (University of Navarra). R. A. thanks Friends of
the University of Navarra for a scholarship.

%
%

\end{document}